\documentclass[prl,twocolumn,aps]{revtex4}
\usepackage{graphicx}

\def\({\left(}
\def\){\right)}

\begin{document}

\title{Scaling in the one-dimensional Anderson localization problem in the
region of fluctuation states.}
\author{L. I. Deych}
\author{M. V. Erementchouk}
\author{A. A. Lisyansky}
\affiliation{Physics Department, Queens College of City University
of New York, Flushing, NY 11367}

\begin{abstract}
We numerically study the distribution function of the conductivity
(transmission) in the one-dimensional tight-binding Anderson model
in the region of fluctuation states. We show that while single
parameter scaling in this region is not valid, the distribution
can still be described within a scaling approach based upon the
ratio of two fundamental quantities, the localization length,
$l_{loc}$, and a new length, $l_s$, related to the integral
density of states. In an intermediate interval of the system's
length $L$, $l_{loc}\ll L\ll l_s$, the variance of the Lyapunov
exponent does not follow the predictions of the central limit
theorem, and may even grow with $L$.
\end{abstract}

\maketitle

\paragraph{Introduction.}
A coherent transport in mesoscopic disordered systems is
characterized by strong fluctuations and non-self-averaging nature
of the transport coefficients such as conductance, $g$, or
transmittance, $T$, \cite{UCFreview,LGP}. Therefore, a description
of the transport in such systems requires dealing with entire
distribution functions of the respective quantities. The scaling
approach to the transport allows one to introduce a reduced
\textquotedblleft macroscopic" description of such distributions
independent of microscopic details of the underlying Hamiltonians
\cite{gang,Anderson} with parameters of the distributions playing
the role of the scaling variables \cite{Anderson}. If the entire
distribution can be parameterized by a single parameter, the
respective \ system is said to obey single parameter scaling
(SPS). A quantity, which is most convenient to work with when
describing the statistics of transport is the Lyapunov exponent
(LE), $\tilde{\gamma}(L)=(1/2L)\ln \left( 1+1/g\right)$, where $L$
is the length of the system \cite{Anderson,Lifshitz}. Finite size
LE, $\tilde{\gamma}(L)$, is self-averaging (approaches a
non-random limit $\gamma $, when $L\rightarrow \infty $
\cite{LGP}), and its distribution approaches a Gaussian form for
asymptotically long systems. The Gaussian distributions are
characterized by two parameters: the mean value, $\gamma $, and
the variance, $\sigma ^{2}$, and the SPS hypothesis suggests that
they are related to each other in a universal way. Such a
relation, which can be expressed in the form
\begin{equation}
\tau =\sigma ^{2}Ll_{loc}=1,  \label{eq:SPS}
\end{equation}%
where $l_{loc}=1/\gamma $ is identified with the localization length, was
first conjectured by Anderson et al.~\cite{Anderson} and reproduced later by
many authors within the framework of the phase randomization hypothesis \cite%
{LGP}. The phase randomization was proven rigorously for in-band
states (those belonging to the spectrum of underlying ordered
systems) for some one-dimensional models (Anderson model
\cite{Goldhirsch1994} and a continuous model with a white-noise
random potential \cite{Titov}) as well as for some
quasi-one-dimensional models \cite{Mirlin1}. For the Lloyd model,
the authors of Ref.~\onlinecite{DeychPRL}  showed that
Eq.~(\ref{eq:SPS}) (corrected by the factor of two) holds for
in-band states even though the distribution of phases is not
uniform.

At the same time, numerical results presented in Ref.~%
\onlinecite{DimaPRL} showed that Eq.~(\ref{eq:SPS}) is not valid
for fluctuation states arising due to disorder outside of the
original spectrum. A boundary between SPS and non-SPS spectral
regions in the exactly solvable Lloyd model was shown to be
determined by a relation $l_{loc}(E)\approx l_{s}(E)$
\cite{DeychPRL}, where $l_{s}$ is a new length, defined through
the number of states, $N(E)$, per unit length, between $E$ and the
closest genuine spectral boundary:
\begin{equation}
l_{s}^{-1}=\sin \left[ \pi N(E)\right] .  \label{eq:ls}
\end{equation}%
In the region of fluctuation states, when $N(E)\ll 1$ or
$1-N(E)\ll 1$, and $l_{s}\gg l_{loc}$, SPS is not valid.
Complimenting analytical
calculations by numerical simulations, the authors of Ref.~%
\onlinecite{DeychPRL} showed that the criterion for SPS found for
the Lloyd model is valid for other models as well.

Thus, it is clear that the problem of the statistics of transport
in the region of fluctuation states requires a separate
consideration. The distinction between this situation and the case
of in-band states can be qualified as a difference between under
barrier tunnelling and over barrier scattering. In 3D this
difference is clear: the latter case corresponds to the spectral
region of extended states with the diffusive transport, while the
former takes place in the region of localized states. The problem
of the under barrier tunnelling in disordered systems was first
considered in \cite{Lifshitz} (for reviews of subsequent papers
see Refs.~\cite{Pollak,Raikh}). In 1D situation all states are
localized, and the transmission for all energies can be described
as a resonant tunnelling via rare transparent configurations
\cite{Azbel}. Therefore, the difference between the
two transport regimes is more subtle and was noticed only recently \cite%
{DimaPRL,DeychPRL}. Correspondingly, while the case of the in-band
states can be considered settled by the SPS theory, the properties
of the distribution function of conductance/transmittance for pure
one-dimensional case of under barrier tunnelling are studied very
little. Besides obvious fundamental importance, an additional
motivation to deal with this problem comes from the development of
photonic band gap materials, in which new type of fluctuation
photonic states is possible \cite{photonic crystals}. These states
form  ``Lifshits tails" in the band-gaps of disordered photonic
structures, and provide a unique opportunity to study resonant
under barrier tunnelling with scattering of light.

The main objective of the present paper is to study numerically
the distribution function of LE in the region of fluctuation
states using ideas of the scaling approach. The main question,
which we seek to answer is the following: \textquotedblleft Is the
distribution function of the LE in the non-SPS region determined
completely by microscopic details of the respective Hamiltonian,
or can it still be described macroscopically in an universal
manner?" We show that the distribution of conductance in this
region, while not completely universal, still demonstrates
surprising scaling properties. In particular, using Monte Carlo
simulations for the one-dimensional tight-binding Anderson model
with diagonal disorder, we find
that for sufficiently long systems the function $\tau $, introduced in Eq.~(%
\ref{eq:SPS}), depends upon a single parameter, $\kappa =l_{loc}/l_{s}$. We
also find strong deviations of the distribution function from the Gaussian
form. However, the third moment, turns out to have the same scaling behavior
as the variance, indicating that despite the deviation of the distribution
function from the Gaussian, it still can be parameterized by only two
parameters, $l_{loc}$, and $l_{s}$.

\paragraph{Model and technical details.}

We consider the tight-binding model with a diagonal disorder, which is
described by the following equations of motion
\begin{equation}  \label{eq:eq_of_motion}
\psi_{n+1} + \psi_{n-1} + (U_n - E)\psi_n = 0,
\end{equation}
where random on-site energies $U_n$ are described by a uniform
probability distribution: $P(U_n) = 1/(2U)$ if $|U_n| < U$, and
$P(U_n) = 0$ otherwise. LE is defined as
\begin{equation}
\gamma (E)=\lim_{N\rightarrow \infty }\frac{1}{N}\log \left\Vert
T_{N}\cdots T_{1}\right\Vert =\lim_{L\rightarrow \infty
}\tilde{\gamma}(L), \label{eq:LE_def}
\end{equation}%
where $T_{k}$ are transfer matrices
\begin{equation}
T_{k}=\left(
\begin{array}{cc}
E-U_{k} & -1 \\
1 & 0\
\end{array}\right).
\label{eq:transfer_mat_def}
\end{equation}
LE is calculated iteratively using Eq.~(\ref{eq:LE_def}) in a
standard way \cite{Kramer}. To investigate the statistics of
$\tilde{\gamma}(L)$ in systems with the finite length $L$, we keep
the length finite and fixed while collecting statistics from about
120,000 realizations. The integral density
of states for each realization was calculated using the phase formalism \cite%
{LGP} and was averaged over all realizations. The resulting value
was used to calculate the length $l_{s}$ according to
Eq.~(\ref{eq:ls}). Studying the dependence of the distribution of
$\tilde{\gamma}$ on $L$ we take care to have $L\gg l_{loc}$ for
all strengths of the disordered potential, $U$, and values of
energy, $E$. However, in the region
of fluctuation states, where $l_{loc}<l_{s}$, it is possible to have $%
l_{s}>L\gg l_{loc}$. In this regime, which does not exist in the SPS region,
the $L$-dependence of the variance may be different from standard behavior
given by the central limit theorem. In order to verify this assumption, we
considered systems with lengths satisfying both $L<l_{s}$, and $L>l_{s}$.
When collecting statistics, we discarded all data corresponding to $%
l_{loc}<5 $, and $l_{s}>1000L$. This way we ensured that our results are not
influenced by non-representative fluctuations, and states localized over
microscopical regions of the sample.

\paragraph{Results.}

\begin{figure}
\hspace{-0.5in}
  \includegraphics[width=3in,angle=-90]{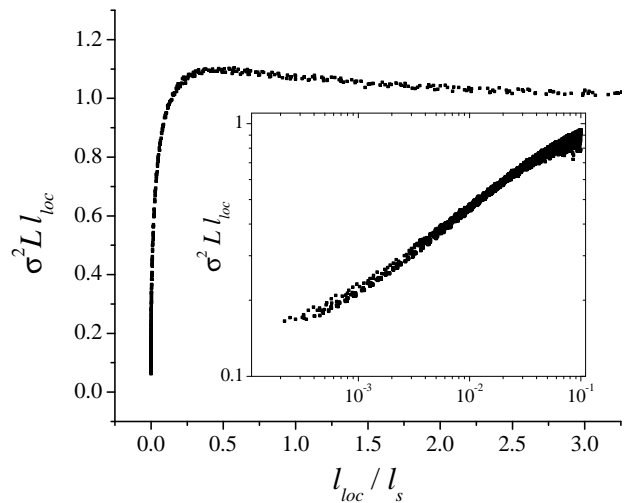}\\
\caption{Dependence of the scaling parameter $\protect\tau$ on $\protect%
\kappa = l_{loc}/l_s$ for a set of potentials ($U = 0.08$ -- $0.155$). On
the insert dependence $\protect\tau(\protect\kappa)$ in the non-SPS region ($%
\protect\kappa \ll 1$) is shown in $\log$-$\log$ scale for $1 <
L/l_s < 5$.} \vspace{-0.3in} \label{fig:scaling_general}
\end{figure}

Fig.~\ref{fig:scaling_general} shows the dependence of the scaling function $%
\tau $ defined in Eq.~(\ref{eq:SPS}) on $\kappa $ in the asymptotic limit of
very long systems, $L\gg l_{s},l_{loc}$. Data used to generate this figure
were obtained for different values of $E$ and $U$, and one can see that they
all fall nicely on the scaling curve $\tau (\kappa )$. For $\kappa >1$, $%
\tau $ approaches its universal SPS value of unity while for
smaller $\kappa$ it steeply decreases. A similar result was
obtained for a periodic-on-average model in Ref.~\cite{DeychPRL},
where the scaling function $\tau (\kappa )$ was originally
proposed. Our results convincingly show that $\sigma ^{2}$ can
indeed be expressed in terms of the scaling function $\tau
(\kappa)$ regardless of the microscopic nature of the model under
consideration.

While the exact shape of the function $\tau (\kappa )$ varies
slightly from model to model (compare to results of
Ref.~\cite{DeychPRL}), the essential qualitative properties of
$\tau $ seem to be quite universal. We are most interested in the
region $\kappa <1$, where $\tau $ demonstrates a sharp decrease.
Analytical calculations carried out in Ref.~\cite{DeychPRL} for
the Lloyd model produced $\tau (\kappa )\approx \kappa $. Our
results show, however, a much steeper decrease of $\tau $. Also,
in the model considered here, $\tau (\kappa )$ must remain
non-zero for $\kappa =0$. Indeed, $\kappa =0$ corresponds to the
exact genuine boundary of the spectrum of our system. Unlike the
Lloyd model, where the spectrum boundary is at infinity, in our
model the boundaries of the spectrum are at $E_{b}=\pm (2+U)$. The
variance of LE does not vanish at finite energies, and therefore
$\tau _{lim}\equiv \tau (0)$ is not equal to zero.

In order to understand the behavior of $\tau$ at $\kappa\ll 1$ we conduct a
detailed study of this region for systems with different $L$. Our results
can be summarized in the following form
\begin{equation}  \label{eq:S_nonSPS}
\tau = C\kappa^{\alpha} + \tau_{lim},
\end{equation}
Replotting $\tau(\kappa)$ in the $log-log$ coordinates for
$\kappa<1$ (see insert in Fig.~\ref{fig:scaling_general}) we see
that while $\kappa$ changes by more than two orders of magnitude,
the data form a good straight line with the exception of points
corresponding to extremely small values of $\kappa$. This means
that $\tau_{lim}$ can be neglected for the most of the non-SPS
region, and becomes significant only in the immediate vicinity of
$E_b$. Using linear regression we can estimate parameters $C$ and
$\alpha$ for systems with different lengths. The results of the
fit reveal that $C$ and $\alpha$ are constants independent of any
parameters of the system under consideration for $L>l_s$. This
result confirms the one-parameter form of $\tau(\kappa)$ given by
the first term of Eq.~(\ref{eq:S_nonSPS}) for sufficiently long
systems. The degree of universality of these coefficients still
remains an open question requiring similar studies of other
models. We can speculate, however, that it is likely that systems
can be divided into several universality classes on the basis of
the values of $C$ and $\alpha$.

\begin{figure}
\hspace{-0.9in}
  \includegraphics[width=3in,angle=-90]{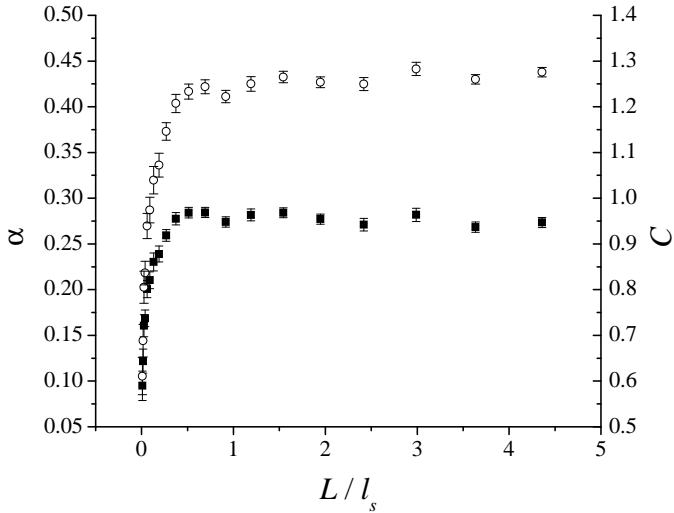}\\
\caption{Index of the scaling parameter $\protect\alpha$ (filled
squares, left axis) and factor $C$ (circles, right axis) as
functions of $L/l_s$.} \vspace{-0.1in} \label{fig:alpha_and_C}
\end{figure}

For shorter systems with $L<l_s$ both $C$ and $\alpha$ show additional
dependence upon the length $L$, see Fig.~\ref{fig:alpha_and_C} where $\alpha$
and $C$ are plotted versus $L/l_s$.

The results of this analysis lead to two important conclusions. First, the
length $l_s$ not only establishes the boundary between SPS and non-SPS
regions of the spectrum, but also determines a crossover system length
marking the transition to systems with a universal single parameter form for
$\tau$. Second, as it was anticipated, in the regime $l_{loc}\ll L<l_s$ the
scaling of the variance of LE changes from the simple $1/L$ dependence to
more a complicated form due to the dependence of $\alpha$ on $L/l_s$. We
attempted to fit this dependence in the region of small $L/l_s$ by several
types of trial functions; the best fit was obtained with $\alpha(L/l_s)\sim
\ln\left(l_s/L\right)$. With this assumption a new scaling for the variance
becomes
\begin{equation}
\sigma^2 \propto \frac{1}{Ll_{loc}}\exp{\left[\alpha(L/l_s)\ln{\kappa}\right]%
}\propto L^{-(1+\ln\kappa)}.
\end{equation}
It is interesting to note that when $\kappa$ decreases, $1+\ln\kappa$
becomes negative and $\sigma^2$ starts growing with $L$ in this interval of
lengths. This behavior can be qualitatively understood from the following
arguments: The condition $L\ll l_s$ means that for the most of the
realizations of the random potential no states exist in the energy interval
under discussion. The transmission through such realizations fluctuates
rather weakly. The greatest contribution to the transmission fluctuations
give those few realizations that can support at least a single state. The
probability for such realizations to arise, grows when the length of the
system increases, resulting in the respective increase of $\sigma^2$.

This behavior, of course, breaks down for very large values of
$l_s$, which correspond to states close to the genuine spectral
boundary, because for these states $\sigma^2$ is determined by a
non-universal correction to $\tau$ given by $\tau_{lim}$. This
limiting value can be found using the weak disorder expansion of
Derrida et al.~\cite{Derrida}, which gives
\begin{equation}  \label{eq:non-universality}
\tau_{lim} \approx \frac{\sigma^2_U}{4 \gamma_0\sinh^2 \gamma_0}\propto
\sqrt{U},
\end{equation}
where $\sigma^2_U = U^2/3$ is the variance of the potential, $\gamma_0$ is
LE in the gap region of the original ordered system, and we use the fact
that $E_b=2+U$ for the system under consideration. One can see from this
expression that $\tau_{lim}$ depends on microscopic characteristics of the
original Hamiltonian. 
However, for a model with the Gaussian distribution of site
energies, the genuine spectral boundary lies at infinity, where
$\gamma_0
\rightarrow\infty $. The first of the equalities in Eq.~(\ref%
{eq:non-universality}), which can be applied to various
distributions, in this case gives
 $\tau_{lim}=0$. We can expect this to be true for all models
with spectral boundaries at infinity. For this class of models, $%
\tau(\kappa) $ gives a completely universal, at least within a given model,
description of the variance of LE.

In order to characterize deviations of the distribution function
of LE from the Gaussian form, we also studied scaling properties
of the third cumulant $\varrho = \langle (\gamma
-\langle\gamma\rangle)^3\rangle$, which describes the skewness of
the distribution function. To analyze scaling properties of
$\varrho$ we introduced a function analogous to $\tau$
\begin{equation}\label{eq:tau_3}
\tau_3 = \varrho L^2 l_{loc}.
\end{equation}

\begin{figure}
\hspace{-0.9in}
  \includegraphics[width=3in,angle=-90]{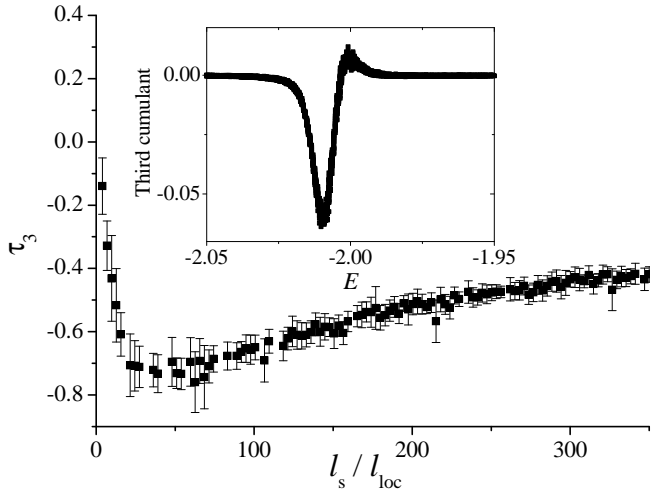}\\
  \caption{Dependence of $\tau_3$ on
$kappa$ deeply inside of the non-SPS region ($\protect\kappa \ll
1$). On the insert scaled third cumulant $\protect\varrho L^2$ is
depicted as a function of the energy near the band edge for the
potential $U = 0.05$ and sample lengths from $1040 $ to
$2160$.}\label{fig:skewness}
\end{figure}

One can see from Fig.~\ref{fig:skewness} that while data for the
parameter $\tau_3$ are rather noisy, it shows a relatively good
scaling behavior as a function of the single parameter $\kappa$ in
the non-SPS region. This fact itself is quite remarkable since it
demonstrates that even deviations from the Gaussian in the region
of fluctuation states can be described within the scaling
procedure suggested here. It is interesting to note that the sign
of the skewness changes not very far from the boundary between SPS
and non-SPS regions of the spectrum. In the SPS region the
absolute value of skewness decreases dramatically becoming
essentially zero within the accuracy of our calculations. This is
illustrated in the insert in Fig.~\ref{fig:skewness}, which
represents the energy dependence of the third moment of the
distribution. The skewness also decreases deeper in non-SPS
regions where it becomes extremely small beyond the genuine
spectral boundary. The quality of our raw data did not allow us to
determine whether $\tau_3$ also depends on $L/l_s$ for shorter
systems, but we expect that $\tau_3$ behaves similar to $\tau$ in
this regard.

\paragraph{Conclusion.}

In this paper we carried out a detail study of the distribution
function of conductance in the spectral region of fluctuation
states. We showed that apart from a small non-universal
contribution, which is only important in the immediate vicinity of
the genuine spectral boundary, the conductance distribution in
this region can be described using a simple scaling approach.
Within this approach the variance is described by the scaling
function $\tau$, Eq.~(\ref{eq:SPS}), and the third moment of the
distribution is characterized by the function $\tau_3$,
Eq.~(\ref{eq:tau_3}). For long enough systems both scaling
functions depend on the single variable $\kappa=l_{loc}/l_s$. The
presence of such a scaling behavior would be natural for the model
with the white noise potential, because such a model has a natural
scaling variable $E/\sigma_U^{3/2}$ \cite{LGP}, and our scaling
parameter $\kappa$ depends upon this only variable
\cite{DeychPRL}. Our numerical results showed, however, that the
parameter $\kappa$ provides a more universal description of the
distribution function valid also outside of the white-noise model.
While we only considered here the tight-binding model, we believe
that our results qualitatively describe statistics of light
transmission through band-gaps of disordered photonic crystals.
Experimental measurements of the transmission distribution in such
systems can be used for verification of our results and as a
method of measuring parameters $l_{loc}$ and $l_s$.

The authors thank Steve Schwarz for reading and commenting on the
manuscript. This work was supported by AFOSR under Contract No.
F49620-02-1-0305, and partially by PSC-CUNY and CUNY collaborative
grants.

\end{document}